\documentclass[reprint, amsmath,amssymb,aps,longbibliography]{revtex4-2}

\usepackage[bottom]{footmisc}% places footnotes at page bottom

\usepackage{kpfonts}
\usepackage[T1]{fontenc}
\usepackage[latin9]{inputenc}
\usepackage{mathrsfs}
\usepackage{bm}
\usepackage{amsmath}
\usepackage{amsthm}
\usepackage{amssymb}
\usepackage{stmaryrd}
\usepackage{mathtools}
\usepackage{changepage}
\usepackage{lineno}
\usepackage{comment}
\usepackage{graphicx}% Include figure files
\usepackage{mathtools}
\usepackage{dcolumn}% Align table columns on decimal point
\usepackage{bm}% bold math
\usepackage{xr}
\DeclarePairedDelimiter\abs{\lvert}{\rvert}%
\makeatletter
\let\oldabs\abs
\def\abs{\@ifstar{\oldabs}{\oldabs*}}

\DeclareFontShape{OMX}{cmex}{m}{b}{<-> cmexb10}{}
\SetSymbolFont{largesymbols}{bold}{OMX}{cmex}{m}{b}

\usepackage{stackengine}
\stackMath

\edef\ordinarycolon{\mathchar\the\mathcode`: }
\edef\ordinaryequals{\mathchar\the\mathcode`= }

\usepackage{breqn}
\AtBeginDocument{%
	\catcode`^=12
}

\allowdisplaybreaks

\let\cat@comma@active\@empty

\usepackage{color}

\usepackage{hyperref}
\usepackage[capitalize]{cleveref}

\newif\ifnotes
\notestrue

\usepackage{color}% for colors
\RequirePackage[dvipsnames,usenames]{xcolor}

\makeatother

\newcommand{\ba}{\begin{eqnarray}}
	\newcommand{\ea}{\end{eqnarray}}

\newcommand{\eq}[1]{\begin{align}#1\end{align}}

\newcommand{\lr}[1]{\langle #1 \rangle}

\newcommand{\III}{{\mathcal{I}}}
\newcommand{\G}{{\mathcal{G}}}

\newcommand{\V}{{\mathcal{V}}}

\newcommand{\PP}{{\mathcal{P}}}
\renewcommand{\G}{{\mathcal{L}}}
\newcommand{\C}{{\mathcal{C}}}

\renewcommand{\L}{{\mathbb{L}}}
\newcommand{\AAA}{{\mathcal{A}}}

\newcommand{\NN}{\mathcal{N}}

\newcommand{\Lu}[1]{\L(p_{x_{#1}}(0),p_{x_{#1}}(\tau))}

\newcommand{\OL}[1]{ \G(#1) \backslash \PP(#1) }

\newcommand{\Anew}[3]{\lr{\AAA^{#1}_{#3}}_{#2}}

\newcommand{\EPRi}[3]{\dot{\zeta}^{#1}_{#3}(#2)}
\newcommand{\EPi}[3]{\zeta^{#1}_{#3}(#2)}

\newcommand{\RML}[3]{K^{#1}_{#2}(#3 t)}
\newcommand{\PFL}[3]{\RML{#1}{#2}{#3} \, p_{#1}(t)}

\usepackage{color}
\definecolor{myblue}{rgb}{.8, .8, 1}

\usepackage{amsmath}
\usepackage{empheq}

\newlength\mytemplen
\newsavebox\mytempbox

\begin{document}
	
	\preprint{APS/123-QED}

	\title{Stochastic thermodynamics of multiple co-evolving systems --- beyond multipartite processes}% 
	
	\author{Farita Tasnim}
	\email[]{farita@mit.edu, web: farita.me}% Your name
	\affiliation{Massachusetts Institute of Technology, Cambridge, MA, USA}
	
	\author{David H. Wolpert}
	\email[]{dhw@santafe.edu, web: davidwolpert.weebly.com}% Your name
	\affiliation{Santa Fe Institute, Santa Fe, NM, USA}
	\affiliation{Complexity Science Hub, Vienna, Austria}
	\affiliation{Arizona State University, Tempe, AZ, USA}
	
	\date{\today}
	
	%TC:ignore
	\begin{abstract}

    Many dynamical systems consist of multiple, co-evolving subsystems (degrees of freedom). These subsystems often depend upon each other in a way that restricts the overall system's dynamics. How does this network of dependencies affect the system's thermodynamics? Prior studies in the stochastic thermodynamics of multipartite processes (MPPs) have approached this question by restricting the system to allow only one subsystem to change state at a time. However, in many real systems, such as chemical reaction networks or electronic circuits, multiple subsystems must change state together. Therefore, studies of MPPs do not apply to such systems. Here, we investigate the thermodynamics of composite processes, in which subsets of subsystems are allowed to change state simultaneously. These subsets correspond to the subsystems that interact with a single mechanism (e.g., a thermal or chemical reservoir) that is coupled to the system. An MPP is simply a (subcase of a) composite process in which all such subsets have cardinality one. We demonstrate the power of the composite systems framework to study the thermodynamics of multiple, co-evolving subsystems. In particular, we derive thermodynamic uncertainty relations for information flows in composite processes. We also derive strengthened speed limits for composite processes. Our results apply to a much broader class of dynamical systems than do results for MPPs, and could guide future studies of the thermodynamics of distributed computational systems.
		
	\end{abstract}
	%\keywords{Suggested keywords}%Use showkeys class option if keyword
	%display desired
	\maketitle
	%TC:endignore
	
%	\listoftodos

Many dynamical systems can be decomposed into a set of multiple co-evolving subsystems. Each subsystem is a degree of freedom. 
For example, information-processing systems such as computers and brains consist of many separate components that evolve together and affect each others' dynamics. 
In practice, such systems are not at thermodynamic equilibrium. 
So, they require energy resources to operate.

Research in the thermodynamics of such distributed computational systems seeks to address how their information-processing capabilities are constrained by the energy resources available to them. 
To conduct such analyses, we must first specify the precise way in which the subsystems co-evolve.  
So far in stochastic thermodynamics, this has often been done by assuming 
the joint dynamics of the subsystems is a multipartite process (MPP)~\cite{horowitz2015multipartiteinfoflow, wolpert2023combining, wolpert_min_ep_2020}. 
% Although a multipartite process does consider the co-evolution of a set of separate subsystems, it additionally assumes 

However, MPPs require that every mechanism (a generalized external system mediating the system's state transitions, e.g., a thermal or chemical reservoir) coupled to the system interacts with only one subsystem.
So, in an MPP, only one subsystem can change state at any given time. 
Unfortunately, at most spatio-temporal scales that we observe, multiple subsystems do change state at the same time. 
As a canonical example, in chemical reaction networks~(\cref{fig:cp_ex}(a)), multiple species counts \textit{must} change state concurrently. 
As another example, the voltages on different conductors in a circuit~(\cref{fig:cp_ex}(b)) must change state at the same time. 

There has been some preliminary work extending the
stochastic thermodynamics of MPPs to address the broader scenario in which each mechanism couples to a set of \textit{multiple} subsystems~\cite{wolpertstrengthenedsecondlaw}. 
Systems with this nature are called \textit{composite systems},
and their dynamics is called a \textit{composite process}. 

Here we extend this preliminary work and obtain new results in the stochastic thermodynamics of composite processes. 
We decompose key quantities (including probability flows, entropy production, and dynamical activity) into contributions from each mechanism. 
We also analyze the network specifying which (set of) subsystems can affect the dynamics of each subsystem.
This network gives rise to \textit{units}, which are subsets of subsystems whose joint dynamics does not depend on the 
state of the rest of the system.
We then use the specification of units and the decomposition of key quantities to derive a wealth of thermodynamic uncertainty relations (TURs).
Finally, we derive a strengthened thermodynamic speed limit theorem (SLT) for composite processes. 
This speed limit provides a tighter restriction on how much the probability distribution over system states can change during a fixed time interval, using the contributions from each mechanism to entropy production and dynamical activity. 
These results also apply to MPPs, since they are a special case of a composite process. 

We begin by reviewing the preliminary work on composite processes, including the specification of units. 
We then present how key quantities decompose into contributions from each mechanism coupled to the system. 
We then present our results for TURs and strengthened SLTs. 
We conclude by discussing our results in the broader contexts of the thermodynamics of constraints and the thermodynamics of computation and by suggesting avenues of future work.

% \dhwc{We should be more detailed in describing our results. In particular,
% given that the special issue is on TURs, we need to describe our TUR results
% in detail.}
% , and we derive strengthened  for composite processes.
% We also analyze special sets of units called ``unit structures'', which allows us to decompose the entropy production of the entire system into an inclusion-exclusion sum of the EP for each unit in the unit structure. 
% We show that this decomposition leads to a wealth of equalities and lower bounds on the global EP. 
% , obtained by substituting existing thermodynamic relations into those for each unit.

\section{Stochastic thermodynamics \\ of composite processes} 

% \dhwc{This first section needs to be broken up. E.g., into a section
% defining composite processes and units, and then a second section,
% describing the thermodynamics of composite systems.}

\subsection{Background on composite processes}

% In this subsection, we review the fundamentals of composite processes as introduced in \cite{wolpertstrengthenedsecondlaw}. 
A composite process is a generalization of
MPPs, describing the co-evolution of a finite set of subsystems, $\NN = \{1, 2, \ldots, N\}$. 
Each subsystem $i$ has a discrete state space $X_i$. 
$x$ indicates a state vector in $X = \bigtimes_{\;i \in \NN} X_i $, the joint state space of the full system.
$x_A$ indicates a state vector in $X = \bigtimes_{\;i \in A} X_i $, the joint state space of the subset $A$. 
The probability that the entire system is in a state $x$ at time $t$ evolves according to a master equation:

\eq{
    \frac{d}{dt} p_x(t) = K^{x'}_{x}(t) p_{x'}(t)
}

\noindent
This stochastic dynamics arises due to couplings of the system with a 
set of mechanisms $\V = \{v_1, v_2, \ldots, v_M \}$. 
In general, each such mechanism $v$ couples to only a subset of the subsystems.  We refer to the set of subsystems to which a mechanism $v$ couples as its \textit{puppet set},
and write it as $\PP(v) \subseteq \NN$. 

As an example, an MPP is a composite process where each mechanism couples to only one subsystem
(although a single subsystem might be coupled to multiple mechanisms~\cite{horowitz2015multipartiteinfoflow}). So in an
MPP, the cardinality of every puppet set is $1$.

%TC:ignore	
\begin{figure}[htbp!]
    \includegraphics[width=0.49\textwidth]{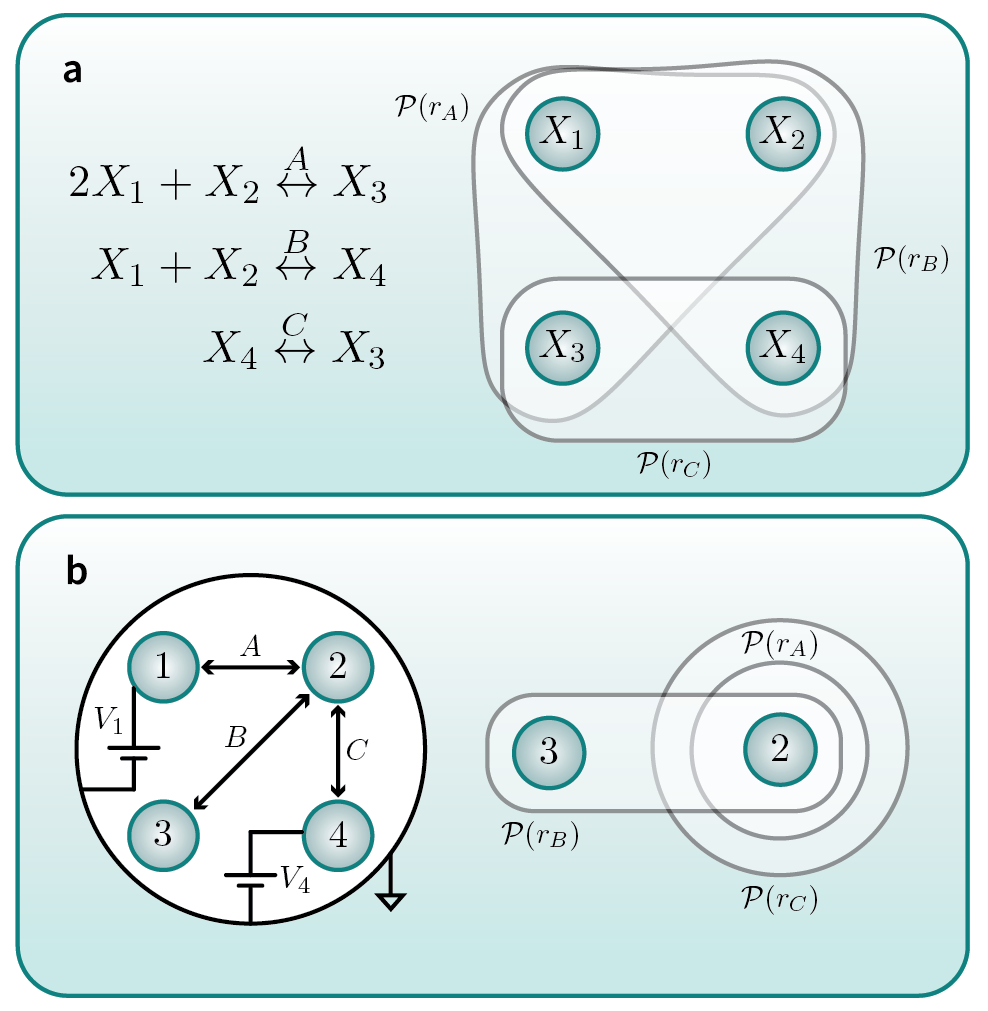}
    \caption{
    Examples of systems whose dynamics can be modeled as composite processes. Each system consists of multiple subsystems (blue circles). Mechanisms are denoted as $r$, and their puppet sets $\PP(r)$ are indicated by translucent white bubbles. 
    % Each black arrow $j \to i$ indicates that the state of subsystem $j$ affects the rate of state transitions in subsystem $i$.
    (a) An example stochastic chemical reaction network consists of four co-evolving species $\{ X_1, X_2, X_3, X_4 \}$ that change state according to three chemical reactions $\{ A, B, C \}$. 
    (b) An example toy circuit consists of four conductors $\{ 1, 2, 3, 4 \}$ that change state via interactions with three devices $\{ A, B, C \}$. 
    } \label{fig:cp_ex}
\end{figure}
%TC:endignore

At any given time, a composite system changes state due to its interaction with at most one mechanism, just as with MPPs. 
Accordingly, the rate matrix of the 
overall system is a sum of
mechanism-specific rate matrices:

\eq{
    K^{x'}_{x}(t) &= \sum_{v \in \V} \delta^{x'_{\NN \backslash \PP(v)}}_{x_{\NN \backslash \PP(v)}} K^{x'_{\PP(v)}, x'_{\NN \backslash \PP(v)}}_{x_{\PP(v)}, x_{\NN \backslash \PP(v)}}(t) \\
    &:= \sum_{v \in \V} K^{x'}_{x}(v;t)
}

\noindent
% which assures that state transitions can occur simultaneously for only the subsystems contained in the puppet set of at most one mechanism at a time 
(Here and throughout, for any two variables, $z, z'$ contained in the same space,
$\delta^{z'}_{z}$ is the Kronecker delta function that equals $1$ when $z' = z$, and equals $0$ otherwise). 
% This generalizes the concept of a multipartite process~\cite{}, in which each mechanism interacts with at most one subsystem, allowing for only one subsystem to change state at any given time. 

We can illustrate composite processes using
a toy stochastic chemical reaction network (\cref{fig:cp_ex}a)~\cite{wachtel2018thermodynamically,rao2016chemrxnPRX, rao2018chemrxn}.
This network involves four co-evolving species $\{ X_1, X_2, X_3, X_4 \}$ that change state according to three chemical reactions $\{ A, B, C \}$ (left). 
The system state is a vector consisting of the number of molecules of each species in the system. Only one reaction can occur at a time, but when a reaction
does occur, \textit{multiple} subsystems all change their state.
For example, in the forward reaction $A$, species $X_1$, $X_2$, and $X_3$ must change state at the same time, by counts of $\{-2, -1, +1\}$, respectively. 
Accordingly, this reaction network is not an MPP. However, it \textit{is}
a composite process.

We can illustrate this composite process in terms of the associated
puppet sets (righthand side of figure).
% the composite process (right) evolving the state of this system of chemical reactions consists of 
There are a total of
three such puppet sets, one for each of the possible chemical reactions.
These three puppet sets are
indicated by translucent bubbles in the righthand
part of the figure.
% Each of these puppet sets corresponds to a mechanism associated with one of the chemical reactions in the system.
The mechanisms of the three puppet sets
are denoted as $r_A$, $r_B$, and $r_C$, and the puppet set of mechanism $r$ is denoted as $\PP(r)$. 
% An arrow $X_i \to X_j$ means that the number of molecules of species $X_i$ affects the rate of transitions in the number of molecules of species $X_j$. 
% \dhwc{I'm not sure what's gained by the arrows? Or by writing $\PP(r)$
% inside the figure? Those clutter things a bit, so if there's no advantage
% to them, I'd suggest removing them.} \dhwc{Also, we should probably
% cite some paper on stochastic thermodynamics of CRNs in the finite
% count limit. E.g., maybe something by Watchel and Esposito.}

As another example, consider a toy electronic circuit (\cref{fig:cp_ex}b)~\cite{freitas2020circuits} consisting of four conductors (the four circles in the left-hand side
of the figure) and three devices (the three bidirectional arrows). 
The state of the system is a vector consisting of the voltage on each conductor. 
Two of the conductors ($1$ and $4$) are ``regulated'', since they are tied directly to fixed voltage sources ($V_1$ and $V_4$). 
The other two conductors ($2$ and $3$) are ``free'' to stochastically change state via the effect of devices $A$, $B$, and $C$. 
% \dhwc{Should we be citing thepaper by Nahuel and Massi?}

The composite process capturing the dynamics
of the state of this circuit is illustrated in the right-hand side
of the figure. There are three puppet sets (each a translucent bubble), each corresponding to a mechanism associated with one of the devices in the system. 
The mechanisms are denoted as $r_A$, $r_B$, and $r_C$, and the puppet set of mechanism $r$ is denoted as $\PP(r)$. 
% % An arrow $i \to j$ means that the voltage on conductor $i$ affects the rate of transitions in the voltage of conductor $j$.  \dhwc{Same comment as above about possibly superfluous graphics inside the figure.}

In an MPP, even though the mechanisms that affect the dynamics of any subsystem $i$ do not affect the dynamics of any other subsystem, in general the dynamics of $i$ will depend on the states of some set of other subsystems. 
For example, in a bipartite process~\cite{horowitz2014bipartiteinfoflow}, both of the subsystems can be modeled as having their own set of mechanisms, but each subsystem's dynamics is governed by the state of the other subsystem as well as its own state. 
% (It is this non-overlap of the puppet sets in bipartite processes that
% is the physical cause of the defining property of bipartite processes, 
% namely that the two subsystems cannot change state simultaneously.)
% (That's how one can
% define the information flow between the two
% subsystems.)

Similarly, in a composite process, the dynamics of each subsystem $i$ can depend on the state of other subsystems in addition to its own state. 
Each such dependency can be represented as an edge in a directed graph.
% We can represent the network coupling all the subsystems to
% their associated leaders as a
% directed graph. 
In the resulting \textit{dependency network} each edge $j \to i$ means that the state of subsystem $j$ affects the rate of state transitions in subsystem $i$. 
We refer to the set of subsystems whose state affects the dynamics of $i$ as the \textit{leaders} of $i$. So
$j \to i$ means that $j$ is a leader of $i$. In any dependency network,
the {leaders} of each subsystem $i$ are its parents, $\text{pa}(i)$.
% , in the dependency network. 
% \dhwc{Should we call it ``dependency network"
% instead of ``dependency network''?}
% \farita{I don't care that much.} \dhwc{Let's go with dependency network then ---
% it's actually \textit{not} a network indicating ``interactions'' in any
% sense. (You and I don't ``interact'' when I watch you, with you not watching me.}

\begin{figure}[htbp!]
    \includegraphics[width=0.4\textwidth]{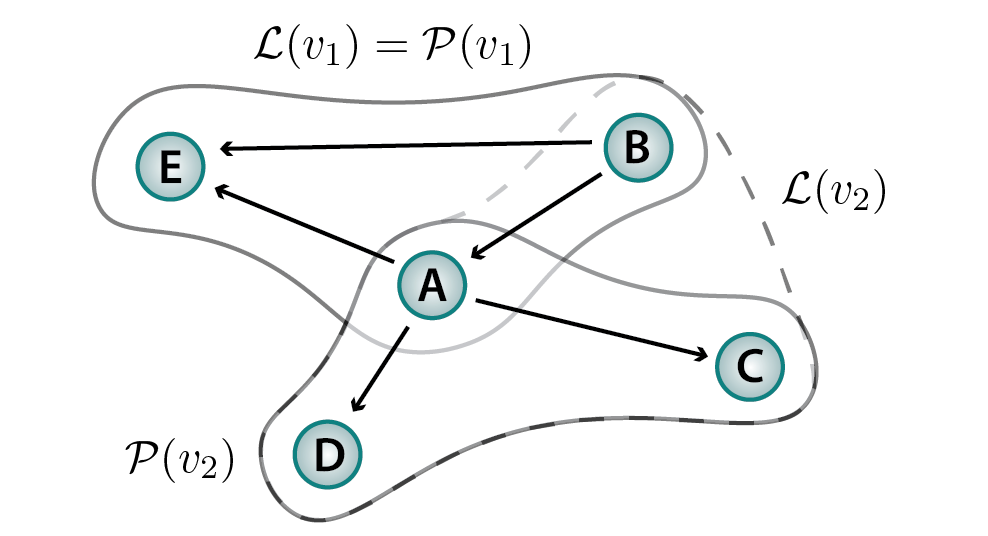}
    \caption{
    The dependency network specifies how the dynamics of each subsystem is governed by the state of other subsystems. This network defines the leader sets in a composite process.
    } \label{fig:leaders}
\end{figure}

The \textit{leader set} for a mechanism $v$ is
defined to be the union of the leaders of each subsystem in the puppet set of $v$: $\G(v) = \bigcup_{i \in \PP(v)} \text{pa}(i)$. 
As an example, even though the puppet set of mechanism $v_2$ in \cref{fig:leaders} is $\{ A, C, D \}$, the leader set of $v_2$ is $\{ A, B, C, D\}$. 

%TC:ignore	
%TC:endignore

\noindent
The leader set of any mechanism is
a (perhaps proper) superset of its puppet
set. 
Accordingly, we can write
% the rates of each particular possible state transition of the
% overall composite system
% is determined by the joint state of the leader set of the mechanism mediating that
% particular state transition: 

\eq{
K^{x'}_{x}(v; t) = 
% \delta^{x'_{\NN \setminus \G(v)} }_{x_{\NN \setminus \G(v)} } \times
    K^{x'_{\G(v)},x'_{\NN \setminus \G(v)} }_{x'_{\OL{v}}, x_{\PP(v)},x'_{\NN \setminus \G(v)} }(v;t)
}
% \dhwc{Note that I expanded this from what was before, to make it be formally
% correct / complete.}

\noindent
With abuse of notation, we can rewrite this
in a way that explicitly embodies the fact that the instantaneous dynamics of the puppet set $\PP(v)$ depends at most on the state of the leader set $\G(v)$, and not on the state of any of the subsystems in $\NN \backslash \G(v)$:

\eq{
 K^{x'_{\G(v)}}_{x'_{\OL{v}}, x_{\PP(v)}}(v;t) := K^{x'}_{x}(v; t)   
}

% \noindent
% which assures that the states of all other subsystems $x_{\NN \backslash \G(v)}$ have no effect on the rates of state transitions in $\PP(v)$. 

% Each mechanism-specific rate matrix can obey local-detailed balance (LDB), so the log-ratio of forward to reverse transitions as mediated by a given mechanism corresponds to a generalized force, which can map to a set of thermal mechanisms, chemical mechanisms, etc.:
% \eq{
%     \ln \frac{K^{x'}_{x}(v; t)}{K^{x}_{x'}(v; t)} &= \mathcal{F}^v_{x' \to x} \\ \label{mechanism-LDB}
%     \spl{
%         &= \beta^v \left( H^v(x',t) - H^v(x,t) \right) \\ 
%         &\quad + \sum_i \mu^v_i \left( N^v(x'_i,t) - N^v(x_i,t) \right) + \ldots
%     }
% }

% \noindent
% This LDB at the mechanism level does not guarantee that LDB holds for the entire system, i.e., that the global rate matrix $K(t)$ also obeys~\cref{mechanism-LDB}. 

% The leader sets resulting from this dependency network gives rise to \textit{units}. 
A \textit{unit} $\omega \subseteq \NN$ is a collection of subsystems such that as the full system's state evolves via a master equation according to $K(t)$, the marginal distribution over the states of the unit also evolves according to its own CTMC:

\eq{
    \frac{d}{dt} p_{x_\omega}(t) = K^{x'_\omega}_{x_\omega}(\omega; t) p_{x'_\omega}(t)
}

\noindent
for some associated rate matrix $K(\omega; t)$. 
% \dhwc{How come this
% $K$ is bold? (It's not bold in the equation just above.)}
% \farita{Matrices are usually bold. The equation above refers to an element from the matrix.} \dhwc{Actually, most usually they aren't written in bold
% font when their entries are unspecified. E.g., the review articles by Esposito, Seifert's tome, information flow paper, etc. A general rule
% of thumb is not to introduce extra notation when it isn't
% needed to clarify a difference that would otherwise be obscure, e.g.,
% between a random variable and the underlying event space, or a trajectory
% over a state space versus an element of that state space.
% But it's your call.}
Intuitively, a unit is any set of subsystems whose evolution is independent of the states of the subsystems outside the unit. 
Typically, a unit is a union of leader sets. 
In such cases no subsystem in the unit has parents outside of the unit. Importantly though, this doesn't prevent
there being a subsystem in the unit that is a leader  
for some subsystem outside of the unit. 
Informally speaking, the boundary of a unit in an dependency network can have outgoing edges, 
even though it cannot have any incoming edges.

Any union of units is a unit, and any non-empty intersection of units is a unit~\cite{wolpertstrengthenedsecondlaw}. 
Note that the entire system $\NN$ itself is a unit. 
% Note also that, in general, the evolution of a subsystem $i \not \in \omega$ may depend on the state of any subsystem $j \in \omega$, even though the reverse is impossible
% by definition. 
We denote the set of all units as $\NN^\dagger$.

Since each separate unit evolves according to its own CTMC, 
all the usual theorems of stochastic thermodynamics
apply to each unit separately. In particular, 
the Second Law~\cite{wolpertstrengthenedsecondlaw} applies, as do the thermodynamic uncertainty relations~\cite{horowitz2017proofFT-TUR, dechant2018MTUR, hasegawa2019FTUR},
the speed limit theorems~\cite{shiraishi_funo_saito_2018, shiraishi2021speed, lee2022speed}, the fluctuation theorems~\cite{rao2018DFT},
first-passage time bounds and 
bounds on stopping times~\cite{gingrich2017FPT, neri2017statistics, neri2019integral}, 
etc. 

We highlight that for any pair of nested units $\omega$ and $\alpha \subseteq \omega$, it is true that~\cite{wolpertstrengthenedsecondlaw, wolpert_min_ep_2020}:

\eq{
    \dot{\sigma}^{\omega}(t) \geq \dot{\sigma}^{\alpha}(t) \label{nested-ep}
}

 A set of units $\NN^*$ is called a \textit{unit structure} if it obeys the following properties~\cite{wolpertstrengthenedsecondlaw}:

\begin{itemize}
    \item The union of the units in the set equals $\NN$. 
    $$\NN^* = \{ \omega_1, \omega_2, \ldots  \}: \bigcup_{\omega \in \NN^\dagger} \omega = \NN$$
    \item The set is closed under intersections of its units.
    $$\forall \, (\omega_1, \omega_2) \in \left( \NN^* \right)^2 = \omega_1 \cap \omega_2 \in \NN^*$$
\end{itemize}

We define an inclusion-exclusion sum of a function $f^\omega$ evaluated on every unit $\omega$ in a unit structure $\NN^*$ as

\eq{
    \widehat{\sum}_{\omega \in \NN^*} f^\omega = \sum_{\omega' \in \NN^*_1} f^{\omega'} - \sum_{\omega'' \in \NN^*_2} f^{\omega''} + \sum_{\omega''' \in \NN^*_3} f^{\omega'''} - \ldots
}

For example, the time-$t$ inclusion-exclusion (or ``in-ex'' for short) information reads

\eq{
    \III^{\NN^*}(t) := \left( \widehat{\sum}_{\omega \in {\NN^*}}  S^\omega(t) \right) - S^{\NN^*}(t) 
}

Using the fact that the heat flow into the unit structure also decomposes into an in-ex sum, we can decompose the global EP incurred during a time period $[0, \tau]$ according to 

\eq{ \label{ep-in-ex}
     \sigma^\NN  =   \widehat{\sum}_{\omega \in {\NN^*}}  \sigma^{\omega} - \Delta \III^{{\NN^*}}
}

\noindent
where $\Delta \III^{{\NN^*}}$ is the change in the in-ex information during the time period $[0, \tau]$. 
For a detailed proof of the in-ex decomposition of the global EP, see \cite{wolpert_min_ep_2020, wolpert2023combining}. 
% As a simple example,
% %of \cref{eq:global_EP_decomp_in_ex}, 
% if there are no overlaps between any units in the unit structure,
% % $\omega_i, \omega_j$, $i \ne j$,
% then expected global EP reduces to

% \eq{
%     \sigma^\NN = \sum_\omega \sigma^\omega - \Delta I\left( X^{\omega_1}; X^{\omega_2};  \ldots \right) 
%     \label{eq:in_ex_decomp}
% }

% \dhwc{If this subsection is just a recapitulation of my earlier
% papers, then it should be moved up to a background section.}

One can use the in-ex sum decomposition of the EP in various ways depending on what degrees of freedom are accessible in the system of interest. 
For example, if one can calculate the mismatch cost~\cite{kolchinsky2017initialdist, kolchinsky2021dependence} $\lambda^\omega$ for each unit in the unit structure, then the in-ex sum can be rewritten:

\eq{
    \sigma^\NN  =   \widehat{\sum}_{\omega \in {\NN^*}}  \lambda^{\omega} + \widehat{\sum}_{\omega \in {\NN^*}}  \xi^{\omega} - \Delta \III^{{\NN^*}}
}

\noindent
where $\xi^\omega = \sigma^\omega - \lambda^\omega$ is the ``residual EP'' due to everything aside from the mismatch cost. 

Additionally, a very large number of lower bounds can be obtained on the global EP by replacing any positive $\sigma^\omega$ (or any such set of them) in the in-ex sum with any lower bound (e.g., TUR, SLT, etc.) on the value of that unit's EP. 
(See~\cite{wolpert2023combining} for examples in the special case of MPPs.)
% \dhwc{\cite{wolpert2023combining} actually doesn't exist. I think 
% references to it should instead be to~\cite{wolpert2023combining}? }

\subsection{Decomposition of thermodynamic and dynamical quantities in composite processes}

% \noindent

Since the entire system $\NN$ is itself a unit, we will write all our results in terms of units for the rest of the paper. 

The rate matrix of each unit $\omega$ in a composite process decomposes into rate matrices from each mechanism whose leader set is a subset of $\omega$:
% \dhwc{Did we mean to say ``leader set'', not ``puppet set''?
% (Else, those sums just below should be over $\PP(v)$, not over $\G(v)$.)}

\eq{
    K^{x'_\omega}_{x_\omega}(\omega; t) &= \sum_{v : \G(v) \subseteq \omega} \delta^{x'_{\omega \backslash \G(v)}}_{x_{\omega \backslash \G(v)}} K^{x'_{\G(v)}, x'_{\omega \backslash \G(v)}}_{x_{\G(v)}, x_{\omega \backslash \G(v)}}(t) \\
    &= \sum_{v : \G(v) \subseteq \omega} K^{x'_\omega}_{x_\omega}(v;t)
}

\noindent
Similarly, we can decompose the EP rate of any unit $\omega$: 
% \dhwc{Again, confusion between puppet and leader sets.}

\eq{
    \dot{\sigma}^{\omega}(t) &= \sum_{ \substack{v : \G(v) \subseteq \omega, \\ x'_\omega, x_\omega \neq x'_\omega} } \PFL{x'_\omega}{x_\omega}{v;} \ln \left[ \frac{\PFL{x'_\omega}{x_\omega}{v;}}{\PFL{x_\omega}{x'_\omega}{v;}} \right] \\
    &= \sum_{v : \G(v) \subseteq \omega} \EPRi{v}{t}{\omega} \label{epv_w}
}

% \dhwc{$\zeta$ undefined --- assuming we mean it to be the summand (which
% as I recall is the case), i.e.,
% the flux times the thermodynamic force, we should say so. }

\noindent
into contributions $\EPRi{v}{t}{\omega}$ from each mechanism whose leader set is a subset of $\omega$. 
In particular, since the entire system is a unit whose state transitions are mediated by every mechanism $v \in \V$, the global EP rate decomposes as $\dot{\sigma}^{\NN}(t) = \sum_{v} \EPRi{v}{t}{\NN}$. 

% \eq{
%     \dot{\sigma}^{\NN}(t) &= \sum_{v, x', x} \PFL{x'}{x}{v;} \ln \left[ \frac{\PFL{x'}{x}{v;}}{\PFL{x}{x'}{v;}} \right] \\
%     &= \sum_{v} \EPRi{v}{t}{\NN}
% }

A unit's dynamical activity also decomposes:

\eq{
    \AAA^\omega(t) = \sum_{ \substack{v : \G(v) \subseteq \omega, \\ x'_\omega, x_\omega \neq x'_\omega} } \PFL{x'_\omega}{x_\omega}{v;} = \sum_{v : \G(v) \subseteq \omega} \AAA(v;t)
}
% \dhwc{Don't we
% need to have the sum be over pairs $x'_\omega, x_\omega\ne x'_\omega$? (This
% is what caused me to fumble with my email about this quantity not being
% positive --- it actually identically equal zero, as written.}
\noindent
Similarly, the entire system's dynamical activity can be decomposed as $\AAA^\NN(t) = \sum_{v} \AAA(v;t)$. 
% \dhwc{I assume there should be an equals sign? }
Note that the dynamics of every pair of nested units $\omega, \alpha \subseteq \omega$ must be consistent with one another~\cite{wolpertstrengthenedsecondlaw},
% \eq{
%     \sum_{ x'_\alpha , x'_{\omega \backslash \alpha}} \PFL{x'_\alpha , x'_{\omega \backslash \alpha}}{x'_\alpha , x'_{\omega \backslash \alpha}}{\omega;} = 
% }
% \noindent 
which means that $\AAA_\alpha(v;t) = \AAA_\omega(v;t) = \AAA(v;t)$ for all $\alpha$ and $\omega$. 
% \dhwc{??? You meant to say, ``for all $\omega, \alpha \subseteq \omega$"?}

We denote the probability flow from $x'_\omega \to x_\omega$ due to mechanism $v$ as $\AAA^{x'_\omega}_{x_\omega}(v;t) = \PFL{x'_\omega}{x_\omega}{v;}$. 
We write the \textit{net} probability current from $x'_\omega \to x_\omega$ due to mechanism $v$ as $J^{x'_\omega}_{x_\omega}(v;t) = \AAA^{x'_\omega}_{x_\omega}(v;t) - \AAA^{x_\omega}_{x'_\omega}(v;t)$. 
The total net probability current from $x'_\omega \to x_\omega$ equals the sum of the probability currents due to each mechanism whose leader set is a subset of the unit $\omega$: 
% \noindent
% where $J^{x'}_x (v;t)$ is the probability flow induced by interactions with mechanism $v$ at time $t$. 

% \noindent
% Similarly, for a unit $\omega$

\eq{
    J^{x'_\omega}_{x_\omega}(t) = \sum_{v : \G(v) \subseteq \omega} J^{x'_\omega}_{x_\omega}(v;t) 
}
% Finally, for conciseness of notation, we will also define the sum of forward ($x' \to x$) and reverse ($x \to x'$) probability currents due to mechanism $v$ as $B^{x'}_{x}(v;t) = A^{x'}_{x}(v;t) + A^{x}_{x'}(v;t)$. 

\noindent
Accordingly, we can decompose the master equation for the unit $\omega$ into probability currents induced by each mechanism:

\eq{
    \frac{d}{dt} p_{x_\omega}(t) = \sum_{ \substack{ v : \G(v) \subseteq \omega, \\ x'_\omega \neq x_\omega } } K^{x'_\omega}_{x_\omega}(v;t) p_{x'_\omega}(t) = \sum_{ \substack{ v : \G(v) \subseteq \omega, \\ x'_\omega } } J^{x'_\omega}_{x_\omega} (v;t)
}
% \dhwc{Can't we just jump directly to this equation? I.e., do we
% need to define activities for a specific state transition, e.g., 
% for subsequent analysis? (I couldn't see whether we do.)}

% Of course, 

% \eq{
%     J^{x'}_{x}(t) = \sum_{v} J^{x'}_{x}(v;t)
% }

% \noindent
% Accordingly, we can decompose the master equation into probability currents induced by each mechanism:

% \eq{
%     \frac{d}{dt} p_{x}(t) = \sum_{v, x'} K^{x'}_{x}(v;t) p_{x'}(t) = \sum_{v, x'} J^{x'}_x (v;t)
% }

% \subsection{Unit Structures}

% \dhwc{This section is too small to be standalone.}

\section{Thermodynamic Uncertainty Relations for Composite Processes}

For any unit $\omega$ that is in an NESS, any linear function of probability currents $\C_\omega$ is a current. It can be divided into contributions from each mechanism:

% \dhwc{$J$ is not defined --- and it's not what's in the summand, which instead
% is something like a (triply-indexed) vector of probability currents, 
% $\{J^{x'_\omega}_{x_\omega} (v)\}$. In fact, in the literature, 
% a ``current'' is taken to mean
% what we're writing as $\dot{\C}_\omega(J)$.}

\eq{
    \dot{\C}_\omega &= \sum_{x'_\omega, x_\omega > x'_\omega} J^{x'_\omega}_{x_\omega} C^{x'_\omega}_{x_\omega} \\
    &= \sum_{\substack{ v : \G(v) \subseteq \omega, \\ x'_\omega, x_\omega  > x'_\omega}} J^{x'_\omega}_{x_\omega} (v) C^{x'_\omega}_{x_\omega} \\
    &= \sum_{v : \G(v) \subseteq \omega} \dot{\C}_\omega(v) \label{current-decomposition}
}

\noindent 
where $C^{x'_\omega}_{x_\omega} = - C_{x'_\omega}^{x_\omega}$ is some anti-symmetric function of state transitions, and we have dropped the time dependence in the steady state. 
% Since the entire system is itself a unit, \cref{current-decomposition} holds for any current functional defined over states of the entire system. 

Importantly, the current contribution from each mechanism $\dot{\C}_\omega(v)$ is itself a current. 
% \dhwc{I don't follow this --- are you mixing together
% probability currents and more general currents in this sentence?}
So all of the thermodynamic uncertainty relations (TURs) hold for the time-integrated version of any such mechanism-specific current. 
In an NESS running for a time period of length $\tau$, this mechanism-specific time-integrated current is $\C_\omega(v) = \tau \dot{\C}_\omega(v)$. 
Additionally, since every unit evolves according to its own CTMC, the TURs hold for each unit. 

For example, the finite-time TUR bounds the precision of any current in a CTMC with respect to its EP rate~\cite{horowitz2017proofFT-TUR, pietzonka2017finitetimeTUR}. 
For a composite process, this holds for any unit and any arbitrary time-integrated current:

\eq{
    \sigma^\omega \geq \frac{2 \lr {\C_\omega}^2} {\text{Var}(\C_\omega}
}

\noindent
Additionally, for any mechanism $v : \G(v) \subseteq \omega$ and any associated current $\C_\omega(v)$,

\eq{
    \sigma^\omega \geq \frac{2 \lr {\C_\omega(v)}^2} {\text{Var}(\C_\omega(v))}
}

The vector-valued TUR following \cite{dechant2018MTUR} holds for a vector $\bm{\dot{\C}}_\omega$ of any set of (potentially mechanism-specific) currents $\{ \dot{\C}_\omega \}$ that are not linearly dependent:
% \begin{equation}
%     \bm{\dot{\C}}^T \mathbf{\Xi}^{-1} \bm{\dot{\C}} \leq \frac{\dot{\sigma}^\NN}{2 \tau}
% \end{equation}

% If the set of (potentially mechanism-specific) currents involve state transitions in only the subsystems within the unit $\omega$, then the following vector-valued TUR also holds:
\begin{equation}
    \bm{\dot{\C}}_\omega^T \: \mathbf{\Xi}_\omega^{-1} \: \bm{\dot{\C}}_\omega \leq \frac{\dot{\sigma}^\omega}{2 \tau}
\end{equation}
where $\mathbf{\Xi}_\omega^{-1}$ is the inverse of the covariance matrix of the associated time-integrated currents $\{ \C_\omega \}$.
% \dhwc{Again, how come bold $\C$ when that's not the case above?
% No extra clarity is introduced --- rather than reader just wonders
% what it is that they're missing.}

% As another example, a recently derived result strengthens the finite-time TUR by accounting for the correlation between an arbitrary observable and a current~\cite{dechant2021improving}. 
% This strengthened TUR can be applied to any pair of observable and current in any unit of a composite process. 

Any of these TURs can be useful to bound the entropy production when one has limited access to the system in the sense that one can measure state transitions i) due only to some subset of the mechanisms influencing the system or state transitions or ii) involving some subset of units in the system. 
% \dhwc{Which TURs do you mean to be calling out? The vector-valued one?}

\subsection{Information Flow TURs}

One important quantity in an MPP is information flow~\cite{horowitz2014bipartiteinfoflow, horowitz2015multipartiteinfoflow, hartich_barato_seifert_2016}. 
% \dhwc{We must say information flow \textit{in an
% NESS} --- it is not a current more generally. Also, we should only point
% out that it's a current (in an NESS) \textit{after} we've defined it.}
Here, we extend the concept of information flow to composite processes.
For any unit $\omega$ in an NESS, a set of subsystems $A \subset \omega$, and a set of subsystems $B \subset \omega$ (for which $A \cap B = \varnothing$), the information flow is the rate of decrease in the conditional entropy of the state of $B$ given the state of $A$, due to state transitions in $A$:
\eq{
    \dot{I}^{A \to B} = \sum_{x'_\omega, x_\omega > x'_\omega} J^{x'_\omega}_{x_\omega} \delta^{x'_{\omega \backslash A}}_{x_{\omega \backslash A}} \ln \frac{p_{x_B|x_A}}{p_{x_B|x'_A}} \label{if}
}

\noindent 
So, when $\omega$ is in an NESS, the information flow is a current 
for which $C^{x'_\omega}_{x_\omega} = C^{x'_{\omega \backslash A}, x'_A}_{x_{\omega \backslash A}, x_A} = \delta^{x'_{\omega \backslash A}}_{x_{\omega \backslash A}} \ln \frac{p_{x_B|x_A}}{p_{x_B|x'_A}}$. 
The contribution to that information flow that is due to interactions of the unit with reservoir $v : \G(v) \subseteq \omega$ is itself an information flow
\eq{
    \dot{I}^{A \to B}(v) = \sum_{x'_\omega, x_\omega > x'_\omega} J^{x'_\omega}_{x_\omega}(v) \delta^{x'_{\omega \backslash A}}_{x_{\omega \backslash A}} \ln \frac{p_{x_B|x_A}}{p_{x_B|x'_A}} \label{ifv}
}

\noindent
Since these information flows are currents, the TURs will apply to them. 
This observation in combination with \cref{nested-ep} suggests that the precision of an information flow is (best) bounded by the reciprocal of the entropy production of the smallest unit which contains $A \cup B$. 

% By \cref{current-decomposition}, this information flow can be divided into contributions from every mechanism whose puppet set is a subset of $A$.

% \eq{
%     \dot{I}^{A \to B} = 
% }

% point out interesting that TUR for correlation of current with observable can be applied for information flow and mutual information

\section{Strengthened Thermodynamic Speed Limits for Composite Processes}

Here we derive a speed limit similar to the one in \cite{lee2022speed}, but for composite processes. This speed limit is tighter than the one presented in that paper. Our analysis will hold for an arbitrary unit $\omega$ 
(which could be the entire system $\NN$ itself):

\eq{
    l_\omega \leq \sum_{v : \G(v) \subseteq \omega} \AAA_\omega^{\text{tot}}(v; \tau) \, f \left( \frac{\zeta^v_\omega (\tau)}{\AAA_\omega^{\text{tot}}(v; \tau)} \right)
}
% \dhwc{How about we first state the main result --- Eq. 44 below --- and
% then go on to derive it? That's common in physics (a ``theorem-proof'' 
% construction without using those words). It would make it easier
% for the reader, who just wants the result.}

\noindent
where the dynamics occurs during the time period $[0, \tau]$. Additionally, $l_\omega$ is the total variation distance between the initial (time-$0$) and final (time-$\tau$) probability distributions over states of the unit $\omega$. $\AAA_\omega^{\text{tot}}(v; \tau)$ is the total time-integrated dynamical activity due to mechanism $v$. $\zeta^v_\omega (\tau)$ is the total contribution to the entropy production of unit $\omega$ due to interactions of $\omega$ with mechanism $v$.

We start by bounding the total variation distance between the initial and final (time-$\tau$) probability distributions over states of the unit $\omega$: 

\eq{
    l_\omega &:= \Lu{\omega} = \frac{1}{2}\sum_{x_\omega} \abs{p_{x_\omega}(\tau) - p_{x_\omega}(0)} \\
    &= \frac{1}{2}\sum_{x_\omega} \left|\int_0^\tau dt \frac{d}{dt} p_{x_\omega}(t)\right| \\
    &\leq \frac{1}{2}\int_0^\tau dt  \sum_{x_\omega} \abs{\frac{d}{dt} p_{x_\omega}(t)} 
\label{l1}
}
% 
% The unit $\omega$  the entire system $\NN$ is itself a unit, so
% even though we will present in the analysis for the case of a specific
% unit everything that follows applies to the entire system as well.
In a composite process, we can further bound the integrand:
\eq{
		\sum_{x_\omega} \abs{\frac{d}{dt} p_{x_\omega} (t)} &=  \sum_{x_\omega} \abs{ \sum_{v : \G(v) \subseteq \omega} \sum_{x'_\omega \neq x_\omega} J^{x'_\omega}_{x_\omega}(v;t) } \\
        &\leq  \sum_{v : \G(v) \subseteq \omega} \sum_{x_\omega, x'_\omega \neq x_\omega} \abs{  J^{x'_\omega}_{x_\omega}(v;t) } \label{step1}
  %       \\
		% &\qquad = \sum_{v,x_{\G(v)}} \sum_{x'_{\G(v)} \neq x_{\G(v)}} \sum_{x_{\omega \backslash \G(v)}} \abs{    J^{x'_{\G(v)}, x_{\omega \backslash \G(v)}}_{x_{\G(v)}, x_{\omega \backslash \G(v)}}(v;t) } 
	}

\noindent
We write the time-$t$ ``conditional probability distribution'' of the forward process, under the counterfactual scenario that the process evolves with coupling only to mechanism $v : \G(v) \subseteq \omega$ as
% , $\bm{W}_\omega(v;t)$, as:
% \dhwc{Why is it written as bold outside of an equation but non-bold inside of
% it? (Why use bold for anything?) There should not be any mention of
% the $W$ matrix in the sentence --- it's confusingly redundant with the 
% equation.}
\eq{
    W^{x'_\omega}_{x_\omega}(v;t) 
    % = P(x'_\omega, x_\omega \, | \,  v, \text{jump}) 
    = \frac{(1 - \delta^{x'_\omega}_{x_\omega}) \, K^{x'_\omega}_{x_\omega}(v;t) \, p_{x'_\omega}(t) }{\AAA_\omega(v;t)}
}

\noindent 
Intuitively, this can be interpreted as a conditional
probability that if a jump occurs at $t$ due reservoir $v : \G(v) \subseteq \omega$, that the
state before the jump was $x'_\omega$ and the state afterwards was $x_\omega$
% \dhwc{This is not a properly defined conditional distribution, for a bunch
% of reasons. Just remove that intermediate part of the definition, and then
% after the equation say, ``Intuitively, this can be interpreted as a conditional
% probability that if a jump occurs at $t$ due reservoir $v$, that the
% state before the jump was $x'_\omega$ and the state afterwards was $x_\omega$''?
% And do we need to say something about }
% \dhwc{Also, don't we need to specify that $\omega$ is in the puppet set
% of $v$? }
We write the same quantity for the reverse process as

\eq{
    \widetilde{W}^{x_\omega'}_{x_\omega}(t)
    % = P(x_\omega, x'_\omega \, | \,  v, \text{jump}) 
    = \frac{(1 - \delta^{x_\omega}_{x'_\omega}) \, K^{x_\omega}_{x'_\omega}(v;t) \, p_{x_\omega}(t) }{\AAA_\omega(v;t)}
}
% \dhwc{Bad notation -- having the asterisk after the $t$. Instead just use
% a tilde above the $W$.}

\noindent
The total variation distance between these matrices $d_{\text{TV}}(\mathbf{W}_\omega(v;t), \mathbf{\widetilde{W}}_\omega(v;t))$ represents how irreversible this counterfactual process (the one driven only by mechanism $v$) is at time $t$. 
Using these definitions, we can rewrite \cref{step1} as 
% \dhwc{We've already introduced notation for total variation distance ---
% should not introduce more notation. Regardless, this is NOT
% total variation distance, since these two $W's$ are not distributions.
% Could you rewrite it? Even once you do though, since the LHS of the following
% equation is positive, how can the equation hold, given that activity
% can be negative?}

\eq{
	\sum_{x_\omega} &\abs{\frac{d}{dt} p_{x_\omega} (t)}
        &\leq 2 \sum_{v : \G(v) \subseteq \omega} \AAA_\omega(v;t) \, d_{\text{TV}}(\mathbf{W}_\omega(v;t), \mathbf{\widetilde{W}}_\omega(v;t)) \label{step2}
	}
% \dhwc{We should give the full derivation of this equation --- it takes six lines
% in Lee et al., even longer if we don't skip steps.}
\noindent
Plugging into \cref{l1}, we obtain

\eq{
    l_\omega \leq \int_0^\tau dt \sum_{v : \G(v) \subseteq \omega} \AAA_\omega(v;t) \, d_{\text{TV}}(\mathbf{W}_\omega(v;t), \mathbf{\widetilde{W}}_\omega(v;t)) \label{step3}
}
% \dhwc{Should we point out that (evidently) Lee et al. assume there is a single reservoir?}

We next make use of the fact that mechanism $v$'s contribution to the EP rate of unit $\omega$ (\cref{epv_w}) can be written in terms of the Kullback-Leibler (KL) divergence between the conditional distributions of the forward and backward
processes as
% between the probability distributions for the forward and reverse counterfactual processes, $D_{\text{KL}}(\mathbf{W}_\omega(v;t), \mathbf{\widetilde{W}}_\omega(v;t))$
% \dhwc{We should derive this from Eq.15 explicitly.}
% \farita{It's literally plug and play from the definitions of zeta and D_KL, and multiplying both sides by the activity. Not sure that deriving it would be a good use of space. I added a reference to equation \cref{epv_w} to make that clear.}
% \dhwc{Sorry for being dense, but as I read \cref{epv_w}, the equality
% would hold if there were \textit{not} the activity term?}
\eq{
    \EPRi{v}{t}{\omega} = \AAA_\omega(v;t) \, D_{\text{KL}}(\mathbf{W}_\omega(v;t), \mathbf{\widetilde{W}}_\omega(v;t)) \label{eprv}
}

\noindent
Any positive monotonic concave function $f$ relates the total variation distance to the KL divergence~\cite{lee2022speed} according to:

% \dhwc{We should give a citation for this.}
\eq{
     d_{\text{TV}}(\bm{p}; \bm{q}) \leq f(D_{\text{KL}}(\bm{p}; \bm{q}))
\label{eq:37}
}

\noindent 
We can use this relationship to relate \cref{eprv} to $l_\omega$. 
Combining \cref{step3,eprv,eq:37},

\eq{
    l_\omega &\leq \int_0^\tau dt \sum_{v : \G(v) \subseteq \omega} \AAA_\omega(v;t) \, f \left( \frac{\EPRi{v}{t}{\omega}}{\AAA_\omega(v;t)} \right)
}

Next define $\zeta^v_\omega = \int_0^\tau dt \EPRi{v}{t}{\omega} $ as the total (ensemble-average) contribution to the EP of unit $\omega$ caused by an interaction of the system with mechanism $v$ during the time period $[0, \tau]$.
Also define $\AAA_\omega^{\text{tot}}(v; \tau) = \int_0^\tau dt \AAA_\omega(v;t)$ as the total (ensemble-average) number of state transitions in the unit $\omega$ that are caused by an interaction of the system with mechanism $v$. 
Then using the positivity of the dynamical activity and of the EP, 
together with the concavity of $f$,
% \dhwc{Don't we need to also use the fact that $f$ is concave?}
we can further bound the right hand side to obtain a general limit for composite processes:

\eq{
    l_\omega \leq \sum_{v : \G(v) \subseteq \omega} \AAA_\omega^{\text{tot}}(v; \tau) \, f \left( \frac{\zeta^v_\omega (\tau)}{\AAA_\omega^{\text{tot}}(v; \tau)} \right) \label{sltcp}
}

% \noindent
This result provides an upper bound on how much $l_\omega$ can change
during the time interval $[0, \tau]$, in terms of the associated activity
of $\omega$ and the contribution of $\omega$ to EP.
So \cref{sltcp} is a thermodynamic speed limit theorem, involving
% just $\omega$. This is our main result. \dhwc{We should be sure
% to describe it in the introduction as such.}

% This equation serves as an effective speed limit, since it bounds the change in the probability distribution over the unit's states using the numbers of mechanism-specific state transitions and mechanism-specific entropy productions incurred during a fixed time period. \dhwc{I don't understand these transitional
% sentences. ???} 
% We can explicitly extract the parameter $\tau$ in the 
% usual way, by using the average frequency of state transitions $ \lr{\AAA_\omega(v)}_{\tau} = \frac{\AAA_\omega^{\text{tot}}(v; \tau)}{\tau}$. 
% However, it suffices to consider a bound on $l_\omega$, so we analyze the speed limit bound as it is written in \cref{sltcp}. 
% The speed limit is tighter when the right hand side of \cref{sltcp} is smaller, which more strongly limits how much the system's distribution over states can change in the same time period. 
% \dhwc{Don't understand that last sentence.}

By comparison, the speed limit in \cite{lee2022speed} applied to a unit $\omega$ reads

\eq{
    l_\omega \leq \AAA_\omega^{\text{tot}}(\tau) \, f \left( \frac{\sigma_\omega(\tau)}{\AAA_\omega^{\text{tot}}(\tau)} \right) \label{sltlee}
}

\noindent
For a composite process, the right hand side of this ``global'' bound expands to

\eq{
    l_\omega \leq \left(\sum_{v : \G(v) \subseteq \omega} \AAA_\omega^{\text{tot}}(v; \tau) \right) \, \, f \left( \frac{\sum_{v : \G(v) \subseteq \omega} \zeta^v_\omega (\tau)}{ \sum_{v : \G(v) \subseteq \omega} \AAA_\omega^{\text{tot}}(v; \tau)} \right)
}

% We note that $\AAA^{\text{tot}}(\tau) = \sum_v \AAA^{\text{tot}}(v; \tau)$ but since $f$ is concave, 

% \eq{
%     f \left( \sum_v \frac{\sigma^v (\tau)}{\AAA^{\text{tot}}(v; \tau)} \right) \leq \sum_v f \left( \frac{\sigma^v (\tau)}{\AAA^{\text{tot}}(v; \tau)} \right)
% }

% \noindent
% At the same time, we know that for two sets of positive numbers with the same cardinality, $\{ a_i \}$ and $\{ b_i \}$:

% \eq{
%     \sum_i a_i b_i \leq \left( \sum_i a_i \right) \, \, \left( \sum_i b_i \right)
% }

\noindent
% Before choosing an appropriate function $f$, we do not know whether the speed limit for composite processes (\cref{sltcp}) is tighter than the speed limit provided by \cite{lee2022speed} (\cref{sltlee}). 
% \cite{lee2022speed} provides some examples of acceptable functions $f$. 
% For example, if we follow Pinsker's inequality and choose $f = \sqrt{\frac{x}{2}}$, then by the Cauchy-Schwarz inequality, the speed limit for composite processes \textit{is} always tighter than the speed limit provided by \cite{lee2022speed}. 

By Jensen's inequality, the speed limit for composite processes (\cref{sltcp}) is always tighter than the speed limit provided by \cite{lee2022speed} (\cref{sltlee}).
For a concave function $f$, a set of numbers a $x_v$ in its domain, and positive weights $a_v$, Jensen's inequality states that

\eq{
    \left( \sum_v a_v \right) f \left( \frac{\sum_v a_v x_v}{\sum_v a_v} \right) \geq \sum_v a_v f(x_v)
}

\noindent
Setting $a_v = \AAA_\omega^{\text{tot}}(v; \tau)$ and $x_v = \frac{\zeta^v_\omega (\tau)}{\AAA_\omega^{\text{tot}}(v; \tau)}$  proves that \cref{sltcp}) is always tighter than \cref{sltlee}. 
Intuitively, this occurs because we're able to define the mechanism-specific contributions to the EP and activity in a composite process.

% \dhwc{Awkward transition --- I cannot remember ever having read the phrase,
% ``As a note".}
% As a note, 
\cite{lee2022speed} provides some examples of acceptable functions $f$. 
For example, if we follow Pinsker's inequality and choose $f = \sqrt{\frac{x}{2}}$, then the speed limit provided by \cite{lee2022speed} collapses to the speed limit derived in~\cite{shiraishi_funo_saito_2018}. 
If we plug in this choice of $f$ to \cref{sltcp}, extract the parameter $\tau$ by using the average frequency of state transitions $ \lr{\AAA_\omega(v)}_{\tau} = \frac{\AAA_\omega^{\text{tot}}(v; \tau)}{\tau}$, and rearrange terms, we obtain

\eq{
        \forall \omega \in \NN^\dagger: \; \tau \geq \frac{(\Lu{\omega})^2}{2 \left( \sum_{v : \G(v) \subseteq \omega} \sqrt{\EPi{v}{\tau}{\omega} \Anew{v}{\tau}{\omega}} \right)^2}
    }

\noindent
the tightest of which is given by: 

    \eq{
        \tau \geq \max_{\omega \in \NN^\dagger} \frac{(\Lu{\omega})^2}{2 \left( \sum_{v : \G(v) \subseteq \omega} \sqrt{\EPi{v}{\tau}{\omega} \Anew{v}{\tau}{}} \right)^2}
    }

\noindent
This particular speed limit tells us that speed of the evolution of the system's probability distribution cannot be greater than the speed of evolution of the distribution over the coordinates of the ``slowest-evolving'' unit.

\section{Discussion} 

Here we have introduced the stochastic thermodynamics of composite processes. 
This work presents a preliminary analysis of how information flows in a composite process are constrained by the entropy productions of units. 
It also demonstrates that bounds on the speed of transforming a system's probability distribution over states can be tightened with knowledge of the contributions to the entropy production and dynamical activity from each mechanism with which the system interacts. 

This work fits into a growing branch of research on the stochastic thermodynamics of constraints. 
One example
of research in this area investigates the effect of constraints on the control protocol (time sequence of rate matrices evolving the probability distribution)~\cite{kolchinsky2020protocol-constraints} There has also been some important work where the ``constraint'' on such a
many-degree-of-freedom classical system is simply that it be some very narrowly 
defined type of system, whose dynamics is specified by many different kinds 
of parameters. For example, there has been analysis of the stochastic thermodynamics of
chemical reaction networks~\cite{rao2016chemrxnPRX, rao2018chemrxn}, of electronic
circuits~\cite{freitas2020circuits, gao2021circuits, wolpert2020circuits}, and of biological copying mechanisms~\cite{poulton2019nonequilibrium}. 
This work analyzes the consequences of a major class of dynamical constraints that arises because many of these systems are most naturally modelled as a set of 
multiple co-evolving subsystems~\cite{dsouza2009structure, horowitz2014bipartiteinfoflow, horowitz2015multipartiteinfoflow, ito2013information, wolpert_bn_2020, wolpertstrengthenedsecondlaw, wolpert2023combining, wolpert_min_ep_2020}. 
In particular, the main constraints on such systems are that only certain subsets of subsystems can simultaneously change state a given time, and the dependencies between subsystems impose restrictions on their joint dynamics. 

% Resource theory and related research~\cite{wilming2016second} also considers 
% issues concerning broadly applicable constraints and thermodynamics, in the context 
% of quantum mechanics. 

There remain many avenues of potential future work, especially in the thermodynamics of computation. 
Many computational processes consist of multiple, co-evolving systems with the broad set of constraints that allow them to be easily modeled as a composite process. 
Research in this direction would first require formalizing the notion of computation in a composite process. 
One such computation, which equates to the identity map, is simply communication (information transmission).
One could extend the recent study on the fundamental thermodynamic costs of communication~\cite{tasnim2023fundamental} to tie Shannon information theory to the stochastic thermodynamics of composite processes. 
More generally, for any given computation, one could analyze the trade-offs between the energy cost required to implement that computation and the performance (accuracy, time, etc.) of a composite process. 
In particular, there could be rich structure in how the properties of the dependency network in a composite process affects these trade-offs.

\section{Acknowledgements}

\noindent	This work was supported by the MIT Media Lab Consortium, Santa Fe Institute, US NSF EAGER Grant CCF-2221345.
% , and FQXi Grant FQXi-RFP-IPW-1912. 
F.T. and D.H.W. thank Tarek Tohme for initial discussions regarding TURs for information flows in multipartite processes. F.T. thanks Nahuel Freitas for discussions regarding how circuits can be modeled as composite processes.

\bibliography{refs.bib}

\end{document}